\documentclass{sf2a-conf2019}
\usepackage{graphicx}
\usepackage{natbib}  

\def\BibTeX{{\rm B\kern-.05em{\sc i\kern-.025em b}\kern-.08em
    T\kern-.1667em\lower.7ex\hbox{E}\kern-.125emX}}
\bibpunct{(}{)}{;}{a}{}{,}  %%%%%%%%%%%%%  A&A bibliography style
\begin{document}

\TitreGlobal{SF2A 2019}

%%-----------------------------------------------------------------
%%      the top matter
%%

\title{Star formation efficiency in low surface brightness regions}

\runningtitle{Star formation efficiency}

\author{F. Combes}\address{Observatoire de Paris, LERMA, Coll\`ege de France, CNRS, PSL Univ., Sorbonne Univ., Paris}

%% IF Author3 has the same affiliation than Author1:
%\author{C.\,E. Author3$^1$}

%% Keep this line, even if the page will be settled afterwards.
\setcounter{page}{237}

%%-----------------------------------------------------------------

\maketitle

%%-----------------------------------------------------------------
%%        The abstract
%% 
%%  Warning!  within the abstract:
%%  - do not use macros. 
%%  - do not use commands like: \cite, \citet, \citep ... etc.

\begin{abstract}
  Low surface brightness regions are found not only in dwarf and ultra-diffuse galaxies, but also on the outer parts of giant spirals,
  or in galaxy extensions (tidal or ram-pressure tails, outflows or jets). Sometimes molecular gas is detected in
  sufficient quantities to allow star formation, but the efficiency is much lower than in disk galaxies.
  This presentation reviews different environments showing low-surface brightness, their gas
  content and surface densities, and their star formation efficiency.  Some interpretations are
  proposed to account for this low efficiency.
\end{abstract}

%% Insert the keywords (to appear in the ADS indexing)
%% Keywords must be separated by a comma

%%-----------------------------------------------------------------

\section{XUV disks}
\begin{figure}[ht!]
 \centering
 \includegraphics[width=0.8\textwidth,clip]{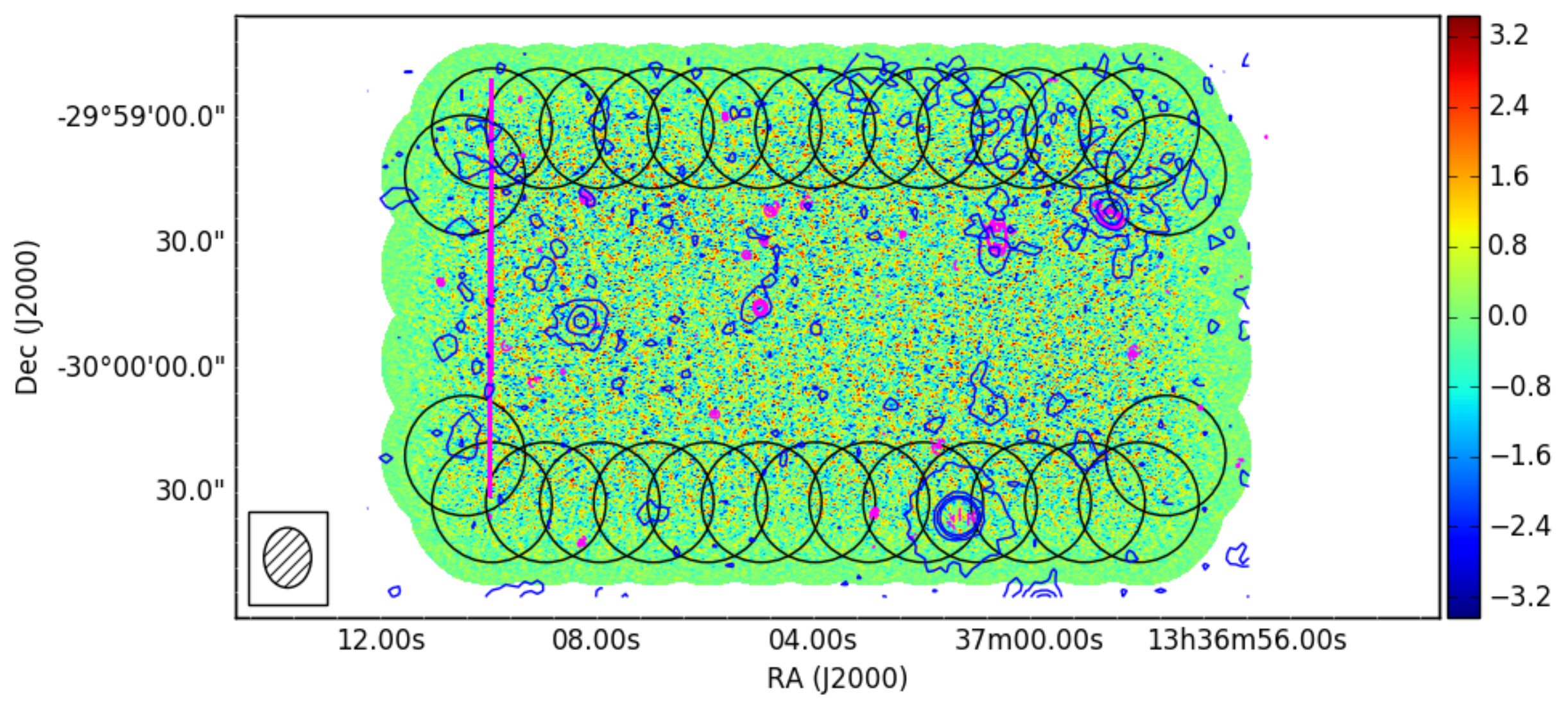}      
  \caption{Schematic representation of the 121-point mosaic of the ALMA observations,
	in the XUV disk of M83.The black circles indicate some of the 27''-diameter
	primary beams of the CO(2-1) emission. In the background, the color map is the
	moment zero of the CO(2-1) data cube. The magenta contours are H$\alpha$ and
	the black contours are FIR 24$\mu$m emission. From \citet{Bicalho2019}.}
  \label{fig1}
\end{figure}

%%---------------------
The extended ultraviolet disks (XUV disks) have been discovered by the GALEX satellite,
and are characterized by UV emission well beyond the optical disk, traced by an H$\alpha$
emission drop. A prototypical example of XUV disks is the galaxy
M83 \citep{Thilker2005}.  To explore the star formation efficiency in the outer
parts of the galaxy, we conducted  ALMA observation in CO(2-1) of the M83 XUV disk,
with a spatial resolution of 17pc x 13pc, well adapted to detect Giant Molecular Clouds (GMC).
Although a significant region was observed (about 2 x 4 kpc) with a 121 point mosaic
(see Figure   \ref{fig1}),
and although the region includes several HII regions and is rich in HI-gas, no CO emission was detected
\citep{Bicalho2019}.
This result is surprising, especially since we detected CO emission in another XUV disk
with the IRAM-30m, e.g. in M63  \citep{Mirka2014}.

\begin{figure}[ht!]
 \centering
 \includegraphics[width=0.5\textwidth,clip]{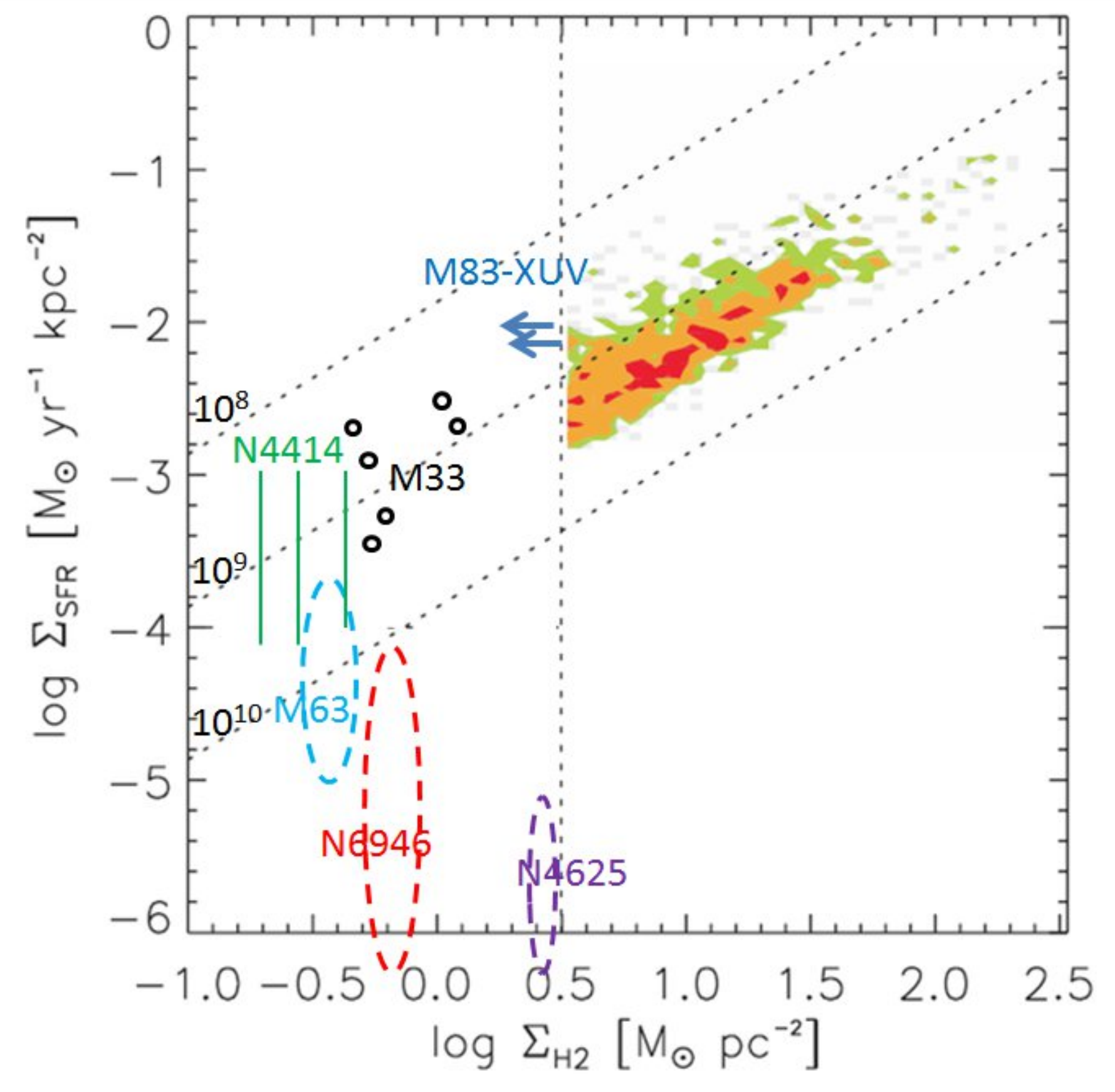}      
  \caption{Resolved Kennicutt-Schmidt diagram, in several galaxies and 
	low surface brighness regions, adapted from \citet{Bigiel2008} 
	and \citet{Verdugo2015}. Dashed ovals indicate the location of 
	XUV disk galaxies. The vertical dash line, at 3 M$_\odot$/pc$^2$
	corresponds to the sensitivity limit of the CO data in \citet{Bigiel2008}.
	Depletion times of 10$^8$, 10$^9$ and 10$^{10}$ years are represented by 3 inclined
	dashed lines. When considering two of the main H$\alpha$ regions, 
	of $\sim$ 200pc size, and the upper limits
	on their molecular gas surface density, we obtain the blue horizontal arrows
	in M83.  From \citet{Bicalho2019}. }
  \label{fig2}
\end{figure}

A compilation of all results from the
literature is plotted in the Kennicutt-Schmidt diagram of Figure  \ref{fig2} ,
in comparison with normal galaxies. This diagram is focussed on the H$_2$ gas, and
shows the depletion time of the local main sequence galaxies of
2x 10$^9$ yrs \citep{Bigiel2008}. The outer disks of M63, NGC~6946 and NGC~4625 have
a much larger depletion time, or equivalently a much lower star formation efficiency (SFE),
by several orders of magnitude.  As for M83, globally the large region observed will
also have a low SFE; however, if we consider the two main HII regions, of size $\sim$ 200pc,
and compute the upper limits of CO emission, the non detection is an exception
(cf Figure  \ref{fig2}).  This absence of CO cannot be attributed solely to a low metallicity,
since the gas abundance in this region is half solar \citep{Bresolin2009}. It is possible that
in these outer regions, molecular clouds are not enough shielded from the UV field,
and the CO molecules are photo-dissociated, while the H$_2$ is still there. Clouds could
be smaller, and the carbon mostly in C and C$^+$.

\section{Cooling filaments}
%%---------------------
Another situation where the SFE is low involves gas flows in cool core clusters.
The prototype is the Perseus cluster, where H$\alpha$ filaments are known since a long
time, most of them excited by shocks, but in some places by star formation  \citep{Canning2014}.
The hot X-ray gas reveals large cavities, sculpted by the central AGN with its radio jets
\citep{Fabian2011}. It is now well established that the AGN feedback moderates
the gas cooling, and cold molecular  gas  is detected around the cavities \citep{Salome2006}.
The gas is still raining  down around the cavities towards  the AGN to fuel it.
CO and H$\alpha$  emissions are well correlated in cool core clusters \citep{Salome2003}.
There is a relation between the star formation rate (SFR) and the gas cooling rate
with a slope larger than unity for strong cooling rates \citep{McDonald2018}.
For low cooling rate, however (lower than 30 M$_\odot$/yr) SFR and cooling rate are not
correlated, pointing to SFR fueled by recycled gas then. The star formation efficiency
may then be low; globally, for the Perseus cluster, it is not far from the mean,
as seen in Figure 3.

\begin{table}
   \centering
        \begin{tabular}{p{7cm}p{9cm}}
 \includegraphics[width=7cm]{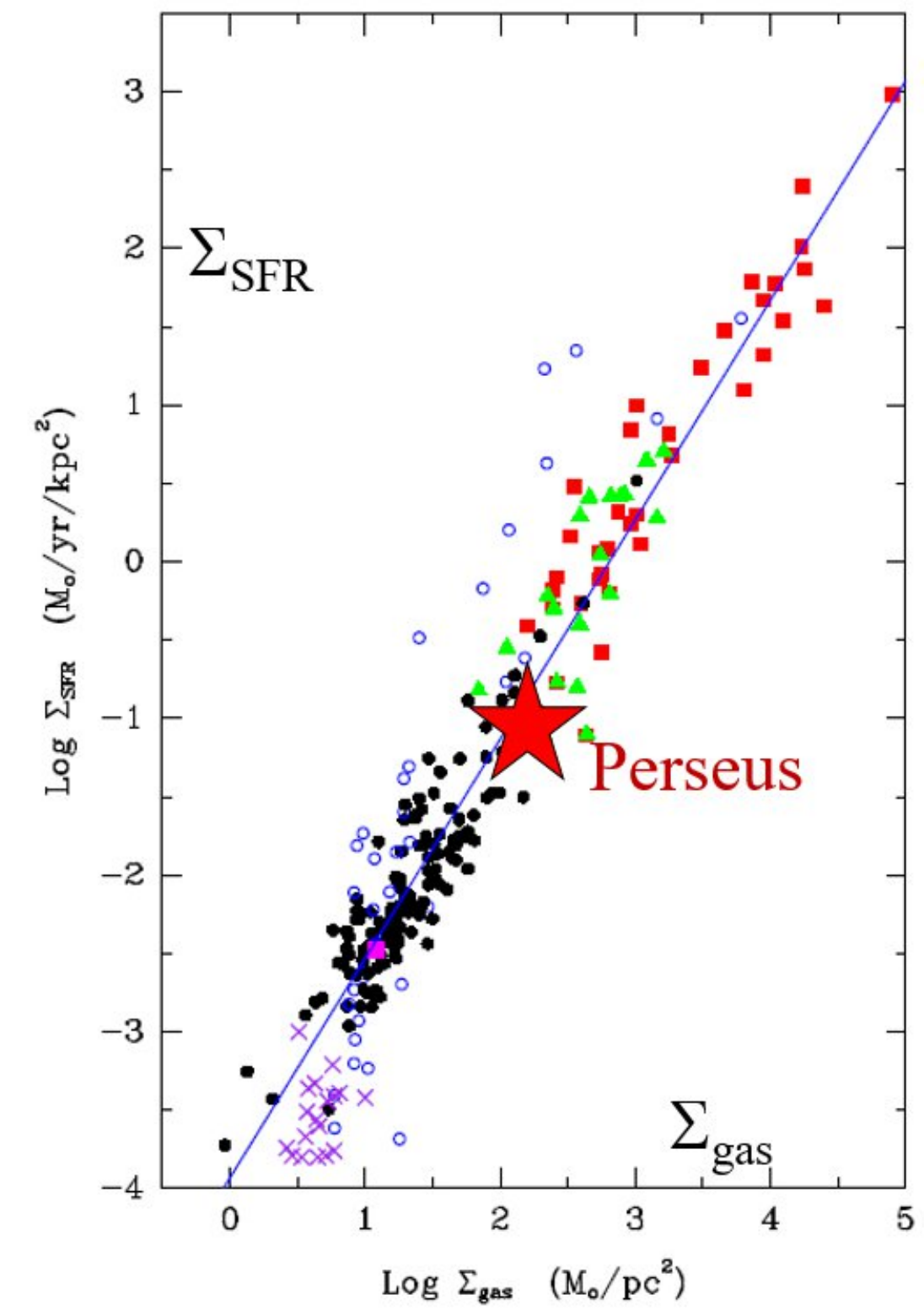} &
 \includegraphics[width=9cm]{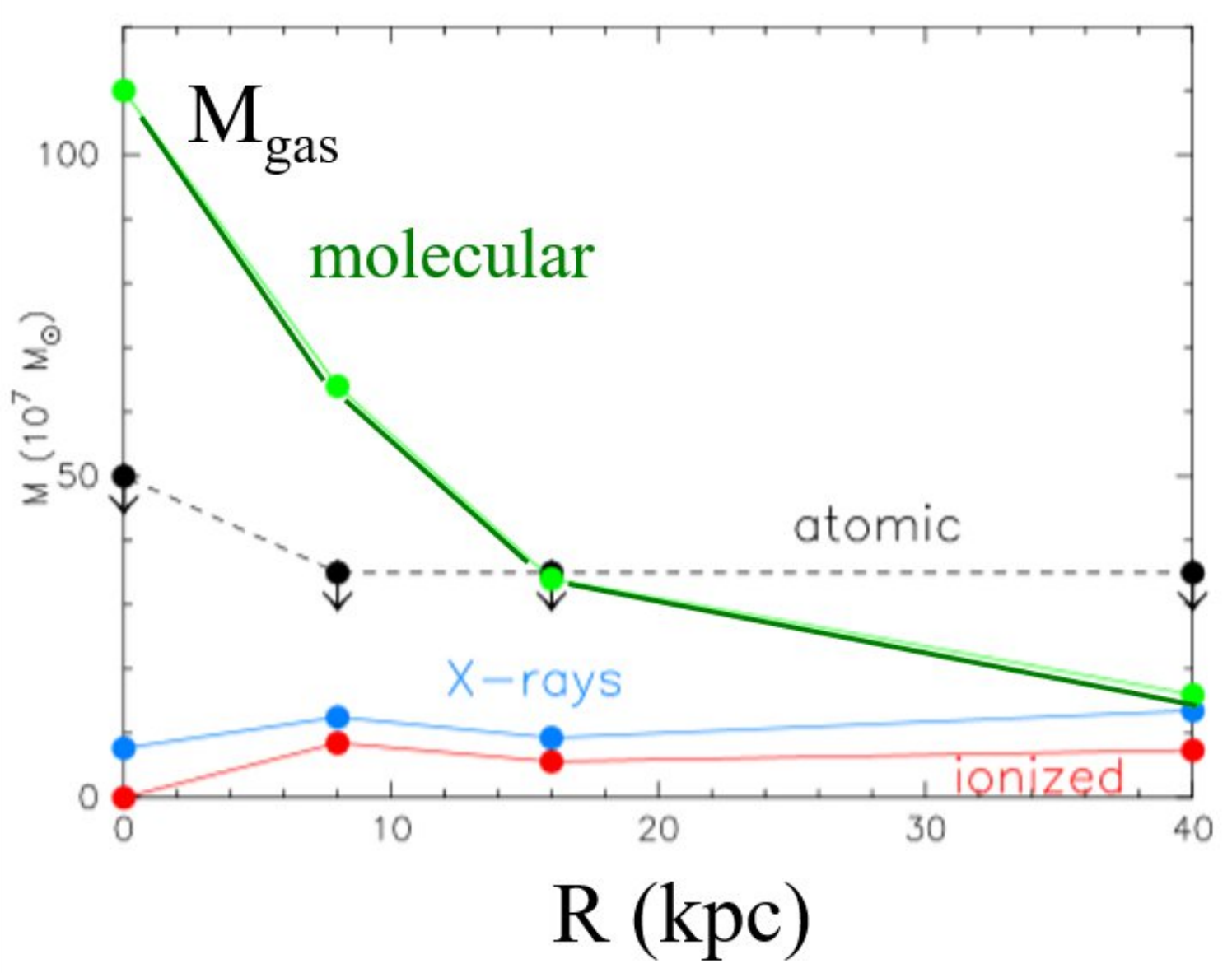}  \\
  {\bf Fig. 3.} The position of Perseus cluster in the global Kennicutt-Schmidt diagram
   is indicated as a large red star, and compared to local main sequence galaxies and
   starbursts  \citep{Kennicutt2012}. &
   {\bf Fig. 4.} Radial distribution of the various gas phases in ESO137-001:
   the molecular gas in green, the atomic HI gas in black (upper limits),
   the X-ray gas in blue, and the H$\alpha$ gas in red.  From \citet{Jachym2014}.    \\
\end{tabular}
 \end{table}

\section{Ram-pressure stripped tails}
%%-----------------------------------------------------
In galaxy clusters, galaxies suffer tidal and ram-pressure stripping
from the Intra-Cluster-Medium (ICM), and lose their gas which constitutes
a diffuse circum-galactic medium. This loss of gas leads to what is called
environmental quenching of star formation. The importance of this quenching
depends on the richness of the cluster. 
Ram pressure is relative mild in the Virgo cluster, where many spirals are
HI deficient, but not perturbed in their molecular gas  \citep{Kenney1989}.

There is however a giant H$\alpha$  tail in the center of Virgo
linking the NGC~4438 galaxy to M86, over scales $\sim$ 100 kpc \citep{Kenney2008}.
Although M86 is a lenticular galaxy devoid of gas, there is CO emission detected
at 10kpc south of this galaxy, in prolongation of the long tail from  NGC~4438. Apparently,
this gas comes from the latter spiral.  It is suprising to find 2 10$^7$ M$_\odot$ of
H$_2$ gas in such an hostile environment,
embedded in hot 10$^7$K ICM gas. Either the H$_2$ molecules have survived during
the 100 Myr  path, or they were re-formed in situ \citep{Dasyra2012}.

One of the H$\alpha$ tail, linking NGC~4388 and M86 is rich in HI gas.
\citet{Verdugo2015} have detected CO emission in some of the clumps along the tail.
When plotted on the Kennicutt-Schmidt diagram, as in Fig. 2, the star formation efficiency
is much lower there than in galaxy disks, similar to  XUV disks \citep{Verdugo2015}.
It is possible that the gas in the tail is more 3D distributed than in 2D-like disks,
and also they are lacking the pressure from the gravity of disk stars.

In richer galaxy clusters, such as the Norma or Coma clusters, ram-pressure can
strip the gas from galaxies much faster: this is the case of ESO137-001 \citep{Jachym2014}.
The tail of 80kpc is spectacular and double in X-ray emission, and in H$\alpha$.
CO emission has been detected easily all along the double-tail, and the molecular gas
dominates the gas content, as shown in Figure 4. The total molecular content of the tail,
a few 10$^9$ M$_\odot$ is larger than the molecular content of the galaxy disk itself.
ALMA has shown that the tail emission is very clumpy, and that the molecular gas is reforming
in situ  \citep{Jachym2019}. The galaxy D100 in Coma has a similar ram-pressure tail,
remarkably straight, and consisting of only one thin component starting from the galaxy center,
as shown by MUSE \citep{Fumagalli2014}. This is expected from a later stage of stripping.
Again the molecular gas dominates in mass the gaseous tail  \citep{Jachym2017},
but the SFE is much lower than in normal galaxy disks.

\section{Importance of pressure}
%%-----------------------------------------------------

In all these low surface density environments, the SFE is found much
lower than in normal disks. This suggests that 
the surface density of stars is very important for the star formation efficiency.
Already \cite{Shi2011} have shown that the SFE in all kinds of environments
is strongly correlated to the stellar surface density, and much less on the gas
surface density.  The HI to H$_2$ transition is favored by external pressure
\citep{Blitz2006}.

An example of radio jet-induced star formation in Centaurus A
have indeed demonstrated that the HI gas is preferentially
transformed in molecular gas on the jet passage \citep{Salome2016}.
Here again the SFE is much lower than normal, the depletion time
in the induced star forming region is between 7 and 16 Gyr.

\section{Conclusions}
%%--------------------

There are several environments showing low surface brightness in stars:
not only dwarf and ultra-diffuse galaxies, but also the outer parts of galaxies, or stars belonging
to circum-galactic regions, either from a cooling flow, or in tidal and ram-pressure
stripped tails. In all these environments, XUV disks, cooling filaments in cool-core clusters,
or ram-pressure tails, the disk pressure due to the gravity
of stars is deficient or non-existing. First, the gas is not confined in a thin
disk, and the star formation might be not only proportional to the gas surface density,
but to the volumic gas density, which is then lower. But also the stellar pressure
is missing to trigger the transformation of diffuse atomic gas to dense molecular gas,
which reduces the efficiency of star formation.
Even in hostile environments like the hot ICM medium, the molecular gas is resilient and can
form in situ, in cooling filaments, in ram-pressure stripped tails. 
Star formation is observed,  but with a low efficiency.

% Optional acknowledgements
% -------------------------
\begin{acknowledgements}
I thank Samuel Boissier for having organised such an interesting workshop.
\end{acknowledgements}

%%-----------------------------
%%   Bibliography
%%-----------------------------
%%
%% The reference list should contain all the references cited in the text, ordered alphabetically by surname (with
%% initials following). If there are several references to the same first author, they should be entered according
%% to the following scheme:
%% 1. One author: chronologically
%% 2. Author, one co-author: alphabetically by co-author, then chronologically
%% 3. Author, two or more co-authors: chronologically.
%%
%% Please note that for papers that have more than five authors, only the first three should be given, followed
%% by "et al."
%%
%% The format for references is the one adopted by A&A (see the example below).
%%
%% To set the reference list in the proper A&A format, we encourage you to use BibTEX and the natbib
%% package instead of the standard 'thebibliography' environment.
%%

%\begin{thebibliography}{}
%\bibitem[Bohr et al.(1992)]{Bohr26} Bohr, N., Einstein, A., \& Fermi, E. 1992, MNRAS, 301, 257

%% The following lines are required when using BibTEX (strongly encouraged!):
\bibliographystyle{aa}  % A&A bibliography style file (aa.bst)
\bibliography{combes_S07} % your references in file: Yourfile.bib

\end{document}